\documentstyle[12pt]{article}
\renewcommand {\c}  {\'{c}}
\newcommand {\cc} {\v{c}}
\newcommand   {\s}  {\v{s}}

\newcommand{\be }{\begin{equation}}
\newcommand{\ee }{\end{equation}}                     

\begin{document}
\baselineskip=24pt
\begin{flushright}
IRB - TH - 6/98
\end{flushright}


\begin{center}
{\bf EXCLUSION STATISTICS, OPERATOR ALGEBRAS AND FOCK SPACE REPRESENTATIONS}
\\[8.0mm]

Stjepan Meljanac $^\dagger$, Marijan Milekovi\c $^\ddagger$ and 
 Marko Stoji\c $^\dagger$ 
\end{center} 

\begin{center}
{\it $^\dagger$  Rudjer Bo\s kovi\c \ Institute , Bijeni\cc ka c.54, 10001 Zagreb,
Croatia}\\
E-mail: meljanac@thphys.irb.hr\\
{\it $^\ddagger$ Prirodoslovno-Matemati\cc ki Fakultet,Zavod za teorijsku
fiziku,\\
Bijeni\cc ka c.32, 10000 Zagreb, Croatia}\\
E-mail: marijan@phy.hr
\end{center}

\setcounter{page}{1}

\vskip 0.2cm            
We study exclusion statistics within the second quantized approach. We consider 
operator algebras with positive definite Fock space and restrict them in a such a way 
that certain state vectors in  Fock space are forbidden ${\it ab \, initio}$.
We describe three characteristic examples of such exclusion, namely exclusion 
on the base space which is characterized by states with specific constraint on 
 quantum numbers belonging to base space 
${\cal M }$ (e.g. Calogero-Sutherland type of exclusion statistics),
exclusion in the single-oscillator Fock space , where some states in 
single oscillator Fock space are forbidden 
(e.g. the Gentile  realization of exclusion statistics) and a combination of these two 
exclusions (e.g. Green's realization of para-Fermi statistics). 
For these types of exclusions we discuss extended Haldane statistics 
parameters $g$, recently  introduced by two of us in 
Mod.Phys.Lett.A 11, 3081 (1996), and 
associated counting rules. Within these three types of exclusions in Fock space 
the original Haldane exclusion statistics cannot be realized.

\bigskip

\newpage

\begin{center}
{\bf Abstract}
\end{center}
We study exclusion statistics within the second quantized approach. We consider 
operator algebras with positive definite Fock space and restrict them in a such a way 
that certain state vectors in  Fock space are forbidden ${\it ab \, initio}$.
We describe three characteristic examples of such exclusion, namely exclusion 
on the base space which is characterized by states with specific constraint on 
 quantum numbers belonging to base space 
${\cal M }$ (e.g. Calogero-Sutherland type of exclusion statistics),
exclusion in the single-oscillator Fock space , where some states in 
single oscillator Fock space are forbidden 
(e.g. the Gentile  realization of exclusion statistics) and a combination of these two 
exclusions (e.g. Green's realization of para-Fermi statistics). 
For these types of exclusions we discuss extended Haldane statistics 
parameters $g$, recently  introduced by two of us in 
Mod.Phys.Lett.A 11, 3081 (1996), and 
associated counting rules. Within these three types of exclusions in Fock space 
the original Haldane exclusion statistics cannot be realized.



\newpage

\section{Introduction}

Statistics plays a fundamental role in the description of macroscopic or 
thermodynamic phenomena in  quantum many-body systems. It is well known that 
the stability of matter, built out of protons and electrons, depends crucially on 
their fermion nature. Also, the Bose - Einstein condensation is essentially 
responsible for the existence of such effects as superfluidity or 
superconductivity. \\
Attempts to generalize these conventional (i.e. Bose and Fermi ) statistics 
date back to Gentile's [1] and Green's [2] works on parastatistics in the 1940s 
and 1950s. Since then a number of papers have been devoted to this topic, 
culminating in  recent years in the  discovery of the fractional quantum Hall 
effect (FQHE)[3], 
theory of anyon superconductivity [4] and the Haldane generalization 
of the Pauli exclusion principle [5].

In principle, there are two distinct approaches to generalized statistics. The starting point 
of the first approach is some symmetry principle, such as
symmetric group (e.g. parastatistics [2]) , braid group 
(e.g. anyon statistics [4]) or quantum groups (e.g. quon statistics [6]). It can  also be
characterized by either an operator algebra of creation and annihilation operators 
with Fock-like representations (second quantization) or  monodromy 
properties of the multiparticle wave function (first quantization).
 For example, Green's parastatistics [2] is based on trilinear 
commutation relations for particle  creation and annihilation operators. For the
para-Bose case of order $p$ exactly those representations of the symmetric group 
$S_N$ with at most $p$ rows in the corresponding Young pattern occur, which means
that at most $p$ particles can be in an antisymmetric state. On the contrary, for 
the para-Fermi case exactly those representations of the $S_N$ with at most $p$ 
columns occur, which means that at most $p$ particles can be in  a symmetric state.
Both cases of parastatistics have some kind of ${\it exclusion}$ built in by their 
very definition, i.e. both a priori exclude certain IRREP's of the symmetric group.

The second approach is based on the state-counting procedure. It is characterized 
by some Hilbert space of quantum states, 
generally without a direct connection with creation and annihilation operators 
acting on Fock-like space. This class includes the  recently   suggested Haldane
 generalization of the Pauli exclusion principle, interpolating between 
Bose and Fermi statistics [5]. The monodromy properties of wave functions are not 
used to define Haldane statistics and the definition of statistics is independent of the 
dimension of space. Instead, 
 the Haldane statistics of a particle [5] is determined by the 
 ${\it statistics \, parameter}$ ${\it g}$ , which for the case of one 
 species of particles is defined as

\be
g=\frac{d_N -d_{N + \Delta N}}{\Delta N },  
\ee

where ${\it N}$ is the number of particles and ${\it d_N}$ is the dimension of 
the one-particle 
Hilbert space obtained by keeping the boundary conditions and quantum numbers of ${\it (N-1)}$
 particles  fixed. For 
bosons, ${\it g}=0$ and for fermions, the Pauli principle implies 
 ${\it g}=1$. \\
Alternatively, Wu has defined the exclusion statistics
by the interpolating counting formula [7], which gives the number of all 
independent N-particle 
states distributed over M quantum states described by M independent oscillators 

\be
D (M,N,g)=\frac{[M + (N-1)(1-g)!]}{N![M-g\, N -(1-g)]!}
\ee

Various aspects of this novel statistics have been investigated [8] and the 
systems exhibiting it 
have been described, including 1D spinons [5] with ${\it g}=\frac{1}{2}$ , 
FQHE quasiparticles [9] and anyonic systems [10] (especially  anyons in 
the lowest Landau level in a 
strong magnetic field). In addition, the Haldane concept of statistics applies
 to the integrable models 
of the Calogero-Sutherland type [11,12]. There is some evidence that it can 
also be helpfull in understanding 
the low-temperature physics of 1D Luttinger liquids [13] and the models which 
exibits the 
Mott metal-insulator transition [14]. Although Haldane statistics is defined 
in an 
arbitrary number of space dimensions,it is evident that  most of these
examples are essentially 1-D systems. 
 Also, there are still some poorly 
understood or unanswered questions such as what is microscopic realization of 
Haldane statistics or its algebraic and group-theoretical characterization.
Furthermore, is there any connection between Green's parastatistics and 
Haldane statistics, since  Green's parastatistics also interpolates 
between bosons and fermions and generalizes the Pauli exclusion principle.\\
There have been several attempts to use operator methods to realize Haldane
exclusion statistics algebraically [15] but the counting rule D(M,N,g), 
calculated for these algebras ( basically of the Gentile type ) differs from the 
Haldane-Wu counting rule. In a previous paper [16] we  found that any operator 
algebra of creation and annihilation operators with a Fock-like 
representations could be described in terms of ${\it extended}$ Haldane 
statistics parameters. Using this result, we described para-Fermi and 
para-Bose statistics as exclusion statistics of the Haldane type and calculated 
a few extended statistics parameters.

In this paper we continue to study exclusion statistics within the second 
quantized approach. 
We consider  various operator algebras with positive definite Fock spaces, 
which 
have the exclusion principle built in them by the very definition and lead to exclusion 
statistics. Since there are many ways to perform such exclusion, 
we describe three characteristic ones, namely exclusion on the base space, 
the Gentile type of exclusion and a combination of these two exclusions. In Section 2
we collect the basic notions of   multimode operator algebras [16,17] which 
are necessary for  Sections 3 - 5, in which we give several examples of 
exclusions mentioned previously. Inspired by exclusion statistics in Calogero-Sutherland model 
[12], 
in Section 3 we define restricted multimode oscillator algebra of quonic type, which depends on 
the 
parameter $\lambda =\frac{p}{q}$ and obeys generalized exclusion principle. We calculate extended 
Haldane statistics parameters and discuss corresponding counting rules for different values of 
$p$ and $q$. We recover Haldane-Wu counting formula for $q = 1$ and integer values of $p$. We also 
briefly mention Fermi and Bose - like exclusion algebras. In Section 4 we discuss Gentile - type 
algebras, with 
common feature that restriction is placed on the single - oscillator Fock space. We find that 
average value of 
extended Haldane statistics parameters and counting rules differs from the original Haldane 
parameters
and counting rules, in agreement with results of Chen et al. in Ref.[15]. In Section 5 we discuss 
Greeen's and Palev's parastatistics as types of exclusion statistics. Finally, in Section 6 we 
briefly 
summarize the main results of the paper.



\section{Definition of the multimode oscillator algebras}
\subsection{Fock space}
\setcounter{equation}{0}
In this section we briefly review the definition of  general multimode 
oscilator algebras   possessing Fock-like representations and well defined 
number operators [16,17].\\
We start with Hermitian conjugated pairs of annihilation and creation operators,
$\{ a_i$,$a^{\dagger}_i|i \in {\cal M } \}$, defined on some base space 
 ${\cal {M}}$.
We build  a Fock-like space starting from the unique vacuum state $|0\rangle$, 
such that 
$\langle 0 |0 \rangle=1,$
$a_i|0\rangle  = 0,\,, \forall i \in  {\cal {M}} $.

An arbitrary multiparticle state can be described as a linear combination of 
monomial 
state vectors $(a^{\dagger}_{i_1} \cdots a^{\dagger}_{i_n} |0\rangle)$ ,and 
the corresponding Fock space ${\cal{F}}_n$ is given as 
\be
{\cal{F}}_n=\{\sum_{i_1\cdots i_n }\lambda_{i_1\cdots i_n }\, a^{\dagger}_{i_1}
\cdots a^{\dagger}_{i_n}
 |0\rangle \; |\;
\lambda_{i_1 \cdots i_n} \in {\bf C }\}.
\ee

The annihilation operators $a_i$ 
act on the space ${\cal{F}}_n$ in such a 
way that

$$
a_ia^{\dagger}_j|0\rangle=\delta_{ij}|0\rangle,
$$
\be
a_ia^{\dagger}_{i_1}a^{\dagger}_{i_2}|0\rangle=\delta_{ii_1}a^{\dagger}_{i_2}|0\rangle +
\Phi^{i}_{i_1 i_2;i_1}\delta_{ii_2}a^{\dagger}_{i_1}|0\rangle ,
\ee
and so on [16,17].

The Fock space (and the corresponding statistics) depends crucially on the 
structure of the base space $\cal{M}$ on which single oscillators are placed.
The simplest base space  ${\cal{M}}$ is the one-dimensional lattice.If the lattice 
is 
finite, we can take ${\cal{M}} = \{ 1,2,\dots M \}$.For infinite lattice we can 
have 
$ \cal{M} = \bf{N} $ or $\bf{Z}$. In the continuum limit, we have $\cal{M} \subseteq 
\bf {R} $. Boundary conditions, being periodic or not, may also be 
 important for statistics. 
Furthermore, one can consider a D-dimensional lattice (finite or infinite) and the 
corresponding continuum limit, with various boundary conditions.
Finally, for the base space one can consider  various manifolds or curved spaces with 
nontrivial topological properties which  may also have important consequences for 
 statistics.

\subsection {Algebra}

We define the $\it {algebra}$ of creation and annihilation 
operators as a normally ordered ( Wick ordered ) expansion $\Gamma_{ij}(a^{\dagger},a)
\equiv  a_ia^{\dagger}_j$ ( no symmetry principle is assumed  ):
\be
\Gamma_{ij} \equiv a_ia^{\dagger}_j=\delta_{ij} + C^{ij}a^{\dagger}_ja_i
+ C^{ij}_{jk,ki}a^{\dagger}_ja^{\dagger}_ka_ka_i+
 C^{ij}_{jk,ik}a^{\dagger}_ja^{\dagger}_ka_ia_k
 $$
 $$
 +C^{ij}_{kj,ik}a^{\dagger}_ka^{\dagger}_ja_ia_k+
 C^{ij}_{kj,ki}a^{\dagger}_ka^{\dagger}_ja_ka_i
 + \cdots ,
\ee
 where C's  are scalar coefficients. Notice that there is no 
need to define any relation between the creation (e.g. $\Gamma_{ij}(a^{\dagger},
a^{\dagger})$) or annihilation (e.g. $\Gamma_{ij}(a,a)$) operators as they appear 
implicitly as norm zero vectors in Fock space. ( For the treatment of the class 
of Wick ordered multimode oscillator
algebras of the form 
$a_ia^{\dagger}_j=\delta_{ij}{\bf 1} + \sum_{k,l}C^{kl}_{ij}a^{\dagger}_la_k$,
see [18].)\\
We also demand that the algebra (2.3) possesses compatible number operators $N_i$ 
 such that $ [N_i,a^{\dagger}_j]=\delta_{ij} a^{\dagger}_i $ and $ [N_i,a_j]=-\delta_{ij}
a_i.$

\subsection{Matrix of inner products ${\cal A}^{(N)}$ and statistics}

For an N-particle state ($a^{\dagger}_{i_{1}}\cdots a^{\dagger}_{i_{N}}|0>$) with fixed indices 
$i_{1},\cdots i_{N}= 1, 2,...M$, there are $\frac{N!}{n_{1}!n_{2}!...n_{M}!}$ (in principle  
different) states obtained by permutations $\pi \in S_{N}$ acting on the state 
$(a^{\dagger}_{i_{1}}\cdots a^{\dagger}_{i_{N}}|0>)$. Here , $n_{1},n_{2}...n_{M}$ are 
eigenvalues of the number operators $ N_i$,
satisfying 
$\sum_{i=1}^{M} n_{i}=N$. From these vectors we form a hermitian matrix 
${\cal A}^{(N)}(i_{1},\cdots i_{N})$ of their scalar products [16,17]. As we have already stated, 
the appearance of null-vectors 
implies corresponding relations between monomials in $a^{\dagger}_{i}$ and reduces the  number of 
linearly 
independent states in $\pi( a^{\dagger}_{i_{1}}\cdots a^{\dagger}_{i_{N}}|0>$). The number of 
linearly
independent states is now given  by the rank of the matrix ${\cal A}^{(N)}$, i.e. $d_{i_{1},\cdots 
i_{N}}={\it rank}{\cal A}(i_{1},\cdots i_{N})$.\\
The set of   $d_{i_{1},\cdots i_{N}}$ for all possible $i_{1},\cdots i_{N}= 1,2,\cdots M $ and all 
integers N  completely characterizes the 
statistics and the thermodynamic properties of a ${\it free}$ system with the corresponding 
Fock space. ( Note that the statistics,i.e. the set 
$d_{i_{1},\cdots i_{N}}$ do not uniquely determine the  algebra given by Eq.(2.3).)

Now, we would like to connect the set $d_{i_{1},\cdots i_{N}}$ with the notion of Haldane 
generalized exclusion statistics . Following Haldane's idea [5] , we define the dimension of the 
one-particle
 subspace of  Fock space
keeping the ${\it (N-1)}$ quantum numbers $i_{1},\cdots i_{N-1}$ inside the N-particle states, 
fixed : 
\be
d_{i_{1},\cdots i_{N-1}}^{(1)}=\sum_{j=1}^{M} d_{j,i_{1},\cdots i_{N-1}}.
\ee
We point out that $d_{i_{1},\cdots i_{N}}$ and $d_{i_{1},\cdots i_{N-1}}^{(1)}$ are integers,
 i.e., no 
fractional dimension is allowed by definition. \\
The number of all independent  N-particle states distributed over M quantum states described by M 
 independent  oscillators ($i=1,2 \cdots M$) is given by 
\be
D(M,N;\Gamma)= \sum_{i_{1},\cdots i_{N}=1}^{M} d_{i_{1},\cdots i_{N}}. 
\ee
Note that $0\leq D(M,N)\leq M^N$ and $D(M,N)$ is always an integer by definition.

The next step is to define the analog of the Haldane statistics 
parameter ${\it g}$.
Recall that Haldane introduced the statistics parameter ${\it g}$ through the change of the 
single-particle 
Hilbert space dimension ${\it d_n}$ ,Eq.(1.1). In a similar way we define 
 ${\it extended }$ Haldane statistics parameters [16] $g_{i_{1},\cdots i_{N-1};j_1\cdots j_k}$
through the change of available one-particle Fock-subspace dimension $d_{i_{1},\cdots 
i_{N-1}}^{(1)}$,
Eq.(2.4),i.e.
\be
g_{i_{1},\cdots i_{N-1};j_1\cdots j_k}=\frac{d_{i_{1},\cdots i_{N-1}}^{(1)} - 
d_{i_{1},\cdots i_{N-1};j_1\cdots j_k}^{(1)}}{k} .
\ee
Note that Eq.(2.6) implies that  extended  Haldane statistics parameters can be 
any rational numbers. Examples of calculation of the matrix  ${\cal A}^{(N)}$ and 
 extended Haldane statistics parameters $g_{i_{1},\cdots i_{N-1};j_1\cdots j_k}$ for 
parastatistics are 
 given in Ref.[16].\\
In the next sections, which constitute the core of the paper, we discuss extended  Haldane 
statistics 
parameters for various types of generalized 
exclusion statistics.




\section{Restricted algebras and projected Fock spaces}

\setcounter{equation}{0}
As we have seen,  one can define extended Haldane 
parameters for any algebra. 
To study exclusion statistics within the second quantized approach, we start 
with an operator algebra and its positive definite Fock space representation. 
Then we restrict the algebra in a such way that certain state vectors in 
 Fock space are forbidden, while the rest of  Fock space remains 
unchanged. There are many ways to perform such exclusions.
 Here we do not pretend to give a complete list of all possible 
exclusions, but  we describe and analyze three main classes of exclusion 
statistics :
\begin{itemize}
\item{(1)} Exclusion on the base space ${\cal M }$ ( example: the Calogero-Sutherland 
type of exclusion statistics) which is characterized by states with a specific constraint on 
positions, momenta or other quantum numbers belonging to the base space (lattice) 
${\cal M }$.
Generally, we can write (no summation over repeated indices)
$$
a_ia^{\dagger}_ja^{\dagger}_k = \Gamma_{ij}\Theta_{jk}a^{\dagger}_k ,
$$
where $\Theta_{jk}$ is 0 or 1 , depending on whether the  simultaneous appearance of $j$, $k$ is,
 respectively,  
forbidden or allowed. The creation operators of the above $\Theta $ - restricted algebra, 
acting on the vacuum $| 0 >$, formally reproduces all the states of the initial 
Fock space (2.1) of the algebra $\Gamma_{ij}$, Eq.(2.3). However, owing to the appearance 
of $\Theta_{jk}$ in the above restricted algebra, monomial states, which do not obey the
$\Theta_{jk}$ - restriction, will have zero norm and effectively disappear
from the Fock space leading to the projected Fock space. This can be easily seen on two particle 
states 
(see, e.g. Examples 3.4 and 3.5).
\item{(2)} Single-oscillator Fock space  restrictions or Gentile-type exclusion 
( examples: Karabali-Nair algebra [15], genons [20]) where  some states in 
single oscillator Fock spaces are forbidden.
 Let $a^{\dagger} a =\phi(N)$, 
$aa^{\dagger}=\phi(N+1)$, $N$ being the number operator with integer eigenvalues 
 $ n \in {\bf N_0} $ and $\phi(n) \geq 0$ [21].\\
Then we can restrict the algebra by  $a^{\dagger} a =\phi(N)\theta(N)$, where 
$\theta(N)$ is 0 or 1, depending on whether the given N - particle (excitation) 
state is forbidden or allowed. The simplest case [22] is $\theta(n)=1, n\leq p$ and 
$\theta(n)=0, n > p$.
\item{(3)} Exclusions  on the base space  ${\cal M }$ and the single Fock space 
simultaneously, i.e. a combination of exclusions of the first and second type 
(examples: Green's and Palev's parastatistics [2,23]).
Generally,
$$
a_ia^{\dagger}_ja^{\dagger}_k = \Gamma_{ij}\Theta_{jk}\Phi(N_a)a^{\dagger}_k ,
$$
where $\Phi(N_a)$ is a  function of the number operators $N_a$ of the a-th particle. 
In  some cases it is not 
needed to project  states  out  of Fock space, since the algebra itself 
incorporates such exclusions, for example $ a_ia^{\dagger}_j= \Gamma_{ij}\Phi(N)$, 
where $\Phi(N)$ is a functional of the total number operator such that 
$\Phi(n) > 0, n \leq p $ and $\Phi(p+1)=0. $
\end{itemize}

\subsection{ Calogero-Sutherland type of fractional statistics}
Let us start with the dynamical C-S model in 1-D, a well known example of exclusion fractional
 statistics [11,12]. The Hamiltonian  for N particles on a ring of length L is 
 given by 
\be
H= -\sum_{i=1}^N \frac{\partial ^2 }{\partial x_i^2 } + \sum_{i \neq j} \frac{2 \lambda (\lambda 
-1)}
{d^2(x_i-x_j)}
\ee
where $ \frac{\hbar ^2}{2m}=1$ and $d(x)=\frac{L}{2\pi } sin(\frac{\pi x}{L})$ and $\lambda \geq 
0$. The spectrum 
of this Hamiltonian is simple and can be expresed in terms of pseudomomenta $k_j$, 
$j=1,2,...N$, in the following form :
\be
E(k_1,...k_N)=\sum_{j=1}^N k_j^2
\ee
where $k_1<k_2 <..< k_N$ and $k_{i+1}-k_i=\kappa \,( \lambda +n_{i+1} )$,$\,$ $n_{i+1}\in {\bf 
N_0}$, 
$ \kappa = \frac{2\pi }{L}$ and 
$ k_1=\kappa\, ( \lambda \frac{N-1}{2} + n_1)$, implying fractional statistics. 
The ground-state energy is for $n_2=n_3=..=n_N=0$ and reads $E(k^0_1,...k^0_N)= \sum_{j=1}^N ( 
k_j^0 )^2 =
\frac {\pi ^2 \lambda ^2 N (N^2 -1)}{3 L^2}$.
 The structure of the spectrum of Calogero model on the 1-D line  in the harmonic potential 
with frequency $\omega $ or in a box is similar to (3.2), with $\kappa $ depending on 
$\omega $ or on the size L of the box .
Note that $\lambda =0$ implies  bosons (on momentum lattice in units of  $\kappa $ ) with
$k_{i+1}-k_i=\kappa n_{i+1}$ and the ground energy 
$E^B_0=0$. The value $\lambda =1 $ implies fermions with $k_{i+1}-k_i=\kappa (1 + n_{i+1})$ and 
ground energy 
$E^F_0>0$.
For both free bosons and fermions we can write the corresponding creation and annihilation 
operators which satisfy Bose and Fermi algebras,respectively.

 We are inspired and motivated by the relation 
$k_{i+1}-k_i=\kappa (\lambda +n_{i+1})$ to construct the algebra of creation and annihilation 
operators 
characterized by $\lambda >0$, $\lambda \in {\bf R^+}$.To do this , we start with the quon algebra 
[6] of 
creation and annihilation operators $a_i, a^{\dagger}_i$ on the real line 
satisfying
\be
a_ia^{\dagger}_j - q a^{\dagger}_ja_i=\delta_{ij}, \qquad i,j \in {\bf R}.
\ee
If $|q|<1$, the corresponding Fock space is positive definite and for a generic 
N-particle state with mutually different indices there are $( N! )$ linearly independent states 
[19].\\
Without loss of generality, we can restrict ourselves to the choice $q=0$ [6,24].
Then the restricted algebra, of type (1) (i.e. exclusion on the base space  ${\cal M }$), 
corresponding to the
algebra $a_ia^{\dagger}_j=\delta_{ij}$, becomes
\be
a_ia^{\dagger}_ja^{\dagger}_k=\delta_{ij}\Theta_{jk}a^{\dagger}_k
\ee
where
$$
 \Theta_{jk}= \left\{ \begin{array}{ll}
1 &\mbox{if $ k-j = \kappa (\lambda +n $)} \\
0 & \mbox{otherwise},
\end{array}
\right. 
$$
with $\kappa >0,\lambda >0,  n \in {\bf N_0} $.
The allowed N-particle states in  Fock space are of the type 
$(a^{\dagger}_{i_1}\cdots a^{\dagger}_{i_N} |0> )$, $i_1\cdots i_N \in {\bf R}$, with  
$i_{\alpha + 1} - i_{\alpha} = \kappa (\lambda +n)$, $n \in {\bf N_0}$ ( all other states are 
null states and hence forbidden). It is obvious that $d_{i_1\cdots i_N}= 1$. If ${\cal M } =
{\bf R}$ (or infinite lattice), then $d^{(1)}_{i_1\cdots i_N}=\infty $ and the extended statistics 
parameters
$g_{i_1\cdots i_N;j_1\cdots j_k}$ are not well defined. However, it is possible to 
define these parameters in the following way. We choose the N-particle state 
$(a^{\dagger}_{i_1}\cdots a^{\dagger}_{i_N} |0> )$ and then take the sufficiently large cut-off 
from the left and from the right, which includes the given N-particle state. So, we obtain 
a finite segment or a finite lattice with M sites, $M>>N$.\\
We shall discuss several cases, depending on values of $\lambda $.The case when 
$\lambda = p/q $ ($p,q \in {\bf N}$) is ${\it rational}$, is relatively simple. 
In this case Ha [12]  suggested normalization of the pseudo-momenta such that the neighbour 
momenta satisfy $i_{\alpha + 1} - i_{\alpha} = p + n_{\alpha} q $, $n_{\alpha} \in 
{\bf N_0}$ and ${\cal M}$ reduces to an infinite lattice, ${\cal M}= {\bf Z}$, with Fermi 
oscillators placed on each site. 
For N-particle state 
$(a^{\dagger}_{i_1}\cdots a^{\dagger}_{i_N} |0>)$ the number of blocked oscillators is
\be
N p +(p - 1) +\sum_{\alpha=1}^{N-1}\Delta_{\alpha}
\ee
where
$$
 \Delta_{\alpha}= \left\{ \begin{array}{ll}
n_{\alpha} q &\mbox{if $ p\neq q, p\neq 1 $} \\
n_{\alpha} ( p -1 ) & \mbox{if $ p = q $},
\end{array}
\right. 
$$
 Specially, if $p=q=1$  
only N-oscillators are blocked. Note that the case $p=q\neq 1$ does not correspond to the 
standard Fermi oscillators, but the group of $p$ oscillators 
 behaves like ordinary Fermi oscillator. 
If $p=1$, $q\neq 1$, then N oscillators are blocked but internal oscillators inside the 
neighbours are strongly correlated.\\
For the closest N-particle states $n_1=n_2=\cdots =n_{N-1}=0$ and  for the finite 
lattice with M sites $\{1,2,\cdots M\}$, we obtain the dimension of the one - particle subspace,
 Eq.(2.5), as
\be
d^{(1)}_N=\Theta(i_1-p) + [\frac{i_1-p}{q}]^{-} + \Theta(M-i_1-N p +1) + 
[\frac{M-i_1-N p +1}{q}]^{-}.
\ee
Then, it is easy to find extended Haldane statistics parameters, Eq.(2.6)
\be
g_{N\rightarrow N+j}=d^{(1)}_N -d^{(1)}_{N+j} = 
\left\{ \begin{array}{cc}
[\frac{i_1-p}{q}]^{-} -[\frac{i_1-2 p}{q}]^{-} & \mbox{if site "j is left"} \\ \\

[\frac{M-i_1-N p +1}{q}]^{-} - [\frac{M-i_1-(N + 1)p +1}{q}]^{-} & \mbox{if site "j is right"}
\end{array}
\right. 
\ee
Hereafter, $[ x ]^{\pm}$ denotes the minimal (+)/ maximal (-) integer greater / smaller than a 
given number $x$, respectively..
We observe that $g_{N\rightarrow N+j}$ depends on $ M $ and $p,q $ as well, and 
 not just on the ratio $\lambda = p/q $ . Moreover, there does not exist a limit value when 
$M\rightarrow \infty$. 
However, one can define the average value of the statistics parameter $\bar{g}$ for $M, M+1, 
\cdots M+q-1$,
 since ${\it g} $ is periodic 
in ${\it M}$ with period ${\it q}$. We assume that ${\it j}$ is always "on the right" and for  
$M>>N p, M>>q$
we find: 
\be
\bar{g}_{N\rightarrow N+j}= \frac{1}{q}\sum_{\alpha=1}^{q} g_{N\rightarrow N+j}(\alpha) = 
\frac{p}{q}=\lambda
\ee
This follows from the identity:
$$
\sum_{i=1}^q [ \frac{M-i}{q} ]^- - [ \frac{M-i-p}{q} ]^- = p,\qquad p \in {\bf N_0}, q \in {\bf 
N}.
$$
Similarly, one can find $g_{i_1,i_2,\cdots i_N;j}$ for arbitrary N-particle states. They depend on 
${\it M, p, q }$
 and $n_1,n_2, \cdots n_{N-1}$. The average value $\bar{g}$ depends generally on ${\it p, q} $ and 
$n_N$. 
 Simple consideration [12] 
  gives, for general ${\it p}$, ${\it q }$ the Haldane statistics parameter $g^{Hald}$

 \be
 g_{N \rightarrow N+1}^{Hald}=\lim_{M\rightarrow \infty} ( \frac{M-N p - (p-1) - 
 \sum_{\alpha=1}^{N-1}n_{\alpha}}{q} 
 - \frac{M- (N+1) p - (p-1) \sum_{\alpha=1}^{N}n_{\alpha}}{q} ) 
 \ee
 $$
 =  \frac{p}{q} + n_N ,
$$
 which is variable. The above consideration gives, for $p=1$, $g^{Hald} = \frac{1}{q}$ 
 and for $p=q$, $g^{Hald} = 1$.
 Hence, only if $p=1$, $\lambda=\frac{1}{q}$, 
the corresponding C-S model has the Haldane statistics parameter $g^{Hald}=\frac{1}{q}$ and 
$\bar{g}=g^{Hald}$. Moreover, two statistical models with the same $g^{Hald}$ (for example, 
$g^{Hald}=1$) 
are not the same. Namely, the counting rule $D(M,N;p,q)$ depends also on both ${\it p}$ and ${\it 
q }$ 
(not  only on the ratio 
$\lambda = p/q $ ):
$$
D(M,N;p,q)=\sum_{i=1}^{M-(N-1) p}\left( \begin{array}{c} 
[  \frac{M -i - p(N-1)}{q}]^- + N - 1 \\ N-1 
\end{array} \right)=
$$
\be
= ( M - p(N-1)  ) \left( \begin{array}{c}
N - 1 + \alpha \\ N-1
\end{array} \right) 
- q (N-1) \left( \begin{array}{c}
N - 1 + \alpha \\ N
\end{array} \right) ,
\ee
where $\alpha = [ \frac{M -1 - p (N-1)}{g} ]^{-}$.
The above equation follows from the identity [25] 
$$
\sum_{i=0}^{\alpha-1} \left( \begin{array}{c}
n+i\\n
\end{array} \right)
= \alpha \left( \begin{array}{c}
n+\alpha\\n
\end{array} \right)
- n \left( \begin{array}{c}
n+\alpha\\n+1
\end{array} \right)
= \left( \begin{array}{c}
n+\alpha\\n+1
\end{array} \right),
$$
with $\alpha \in {\bf N}$ and $n \in {\bf N_0}$.\\
If $p=q$,
\be
D(M,N;q,q)= M \left( \begin{array}{c}
N-1+\alpha\\N-1
\end{array} \right)
- q(N-1)\left( \begin{array}{c}
N+\alpha\\N
\end{array} \right)
\ee
with $\alpha = [ \frac{M -1}{q}]^- -N+1 $.\\
For $p=q=1$, $D(M,N;1,1)=\left( \begin{array}{c}
M\\N
\end{array} \right).$\\
If $q=1$, $p\in {\bf N_0}$
\be
D(M,N;p,1)=\left( \begin{array}{c}
M+(1-p)(N-1)\\N
\end{array} \right) ,
\ee
and only if $q=1$, Eq. (3.10) coincides with an ad hoc interpolation formula by Haldane and Wu 
[5].
The simple interpolation of the above equation is obtained by $\Theta_{jk}=1$ if $k-j \geq p$ and 
$\Theta_{jk}=0$, $k-j < p \in {\bf N}$, i.e. that a single particle blocks $p$-units (for 
fermions, $p=1$).
 We point out that the case $p=0$, $q=1$  makes sense and reproduces the Bose statistics 
 $D(M,N;0,1)=\left( \begin{array}{c}
 M+N-1\\N
\end{array} \right)$, but the case $p=0$ and $q\neq 1$  corresponds to a generalized Bose 
statistics since
\be
 D(M,N;0,q)= M \left( \begin{array}{c}
N-1+[ \frac{M-1}{q} ]^-\\N-1
\end{array} \right)
- q (N-1) \left( \begin{array}{c}
N-1+[ \frac{M-1}{q} ]^-\\N
\end{array} \right).
\ee
We note that for fractional values of $\lambda =\frac{p}{q}$, the counting rule, Eq.(3.10), is 
completely different, 
even asimptotically, from the Haldane-Wu formula .\\
If the coupling constant $\lambda$ is an ${\it irrational }$ positive number,  then the Ha lattice 
construction 
and the counting formula , Eq.(3.10), cannot be applied. In this case it is more appropriate 
to define
\be
g_{n\rightarrow n+1}=\lim_{M\rightarrow \infty} \{ ( M- (n-1)\, \lambda ) - ( M- n \lambda ) \} 
=\lambda,
\ee
where $\lambda $ is the occupation width of one-particle state, and the counting rule
\be
 D(M,N;\lambda)=\left( \begin{array}{c}
 M+N-1-[ (N-1)\lambda ]^{+}\\N-1
 \end{array} \right).
 \ee
 Here we have assumed that the first particle can occupy M-states and that the whole N-particle 
state is 
 smaller than M. It is interesting to note that if $\lambda \in {\bf N_0}$, the last equation 
coincides 
 with Eq. (3.12) and with the Haldane-Wu formula. However, if $\lambda$ is not an integer, then  
 Eq. (3.15) differs from both  Eq.(3.12) and Haldane-Wu formula.  Eq.(3.14) has the advantage to 
be well 
 defined for any real $\lambda \geq 0$ and if $\lambda = \frac{p}{q}$, then D(M,N) depends only on 
 $\lambda $. 
  In this case, one can define an effective parameter $\lambda_{eff}=\frac{[ (N-1)\lambda ]^{+}}
 {N-1}$.
 
 ${\it Remark}$\\
 The case $p=q=1$ ($p=0,q=1 $) corresponds to non-standard fermions (bosons) since the operators 
 $a^{\dagger}_i$ do not satisfy the commutation relations for ordinary fermions ( bosons ), 
although the statistical 
 properties are the same as for ordinary fermions ( bosons ) with $g=1$ ( $g=0$ ).
 
 In the following subsections we briefly mention possible generalizations of exclusion fractional 
statistics 
 by constructing projected Fock spaces.
 
\subsection{ Fermi-like exclusion statistics}
Let us define a monotonic series (or a finite set)
$$
X=\{ x_n | 0 < x_1 < x_2 \cdots < x_n < x_{n+1} \cdots \}.
$$ 
Then, we easily generalize the condition, Eq.(3.4), to 
$$
 \Theta_{jk}= \left\{ \begin{array}{ll}
1 &\mbox{if $ k-j \in X $}\\
0 & \mbox{otherwise}.
\end{array}
\right. 
$$
This restriction leads to the N-particle states
$$
a^{\dagger}_{i_1}\cdots a^{\dagger}_{i_N} |0>,\qquad i_{\alpha + 1} - i_{\alpha} \in X.
$$
(The energy dependence on $i_{\alpha}$ i.e. the dispersion relation is not specified). All other 
states are 
null-states. The meaning of these restrictions is that only ordered states survive and that 
distances 
between neighbours are "quantized" according to the rule $i_{\alpha + 1} - i_{\alpha} \in X$. This 
rule
generalizes the Pauli exclusion principle and we call the corresponding statistics X-type 
restricted Fermi 
statistics. The statistics parameters $g$ and the counting rules D(M,N;X,F) can be found using 
results of 
Ref.[25]. The special case of this statistics is the C-S type of fractional statistics (Section 
3.1).

\subsection{ Bose-like exclusion statistics}

This is in principle the same kind of exclusions as in subsection (3.2), but with the only 
difference that 
in a given state there may be an arbitrarily large number of particles (excitations), i.e.  the 
$\Theta $ projector 
satisfies
$$
 \Theta_{jk}= \left\{ \begin{array}{ll}
1 &\mbox{if $ k-j \in X \bigcup \{0 \}$}\\
0 & \mbox{otherwise}
\end{array}
\right. 
$$
The corresponding allowed states are 
$$
(a^{\dagger}_{i_1})^{n_1}\cdots (a^{\dagger}_{i_N})^{n_N} |0>, \quad n_1, \cdots n_N \in {\bf N},
\qquad i_{\alpha + 1} - i_{\alpha} \in X.
$$
All other states are null-states. The above restrictions lead to the X-restricted Bose statistics. 
The counting rule 
for N-particle states defined on M-neighbouring sites is 
\be
D(M,N;X,B) = \sum_{k=1}^N \left( \begin{array}{c}
N-1\\k-1
\end{array} \right)
D(M,k;X,F)
\ee
where D(M,k;X,F) is the counting rule for the X-restricted Fermi statistics ( subsection (3.2) ). 
The factor $
\left( \begin{array}{c}
N-1\\k-1
\end{array} \right)
$ follows from the identity after Eq.(3.10).\\
Both  X-restricted Fermi and Bose statistics are examples of permutation non-invariant statistics. 
The special case
$X={\bf N}$ reproduces Fermi (Bose) statistics but the algebra of creation and annihilation 
operators 
differs from the ordinary Fermi (Bose) algebra. 

\subsection{ Restricted Fermi algebra}

 One can start from the permutation invariant Fermi algebra $a_i  a^{\dagger}_{j}= \delta_{ij} -
 a^{\dagger}_{j}a_i $ and restrict it in different ways. For example: 
 \be
 a_i  a^{\dagger}_{j}a^{\dagger}_{k}= (\delta_{ij} - a^{\dagger}_{j}a_i) 
\Theta_{jk}a^{\dagger}_{k},
\ee
where
$$
\Theta_{jk}= \left\{ \begin{array}{ll}
1 &\mbox{if $ |k-j| \geq p$}\\
0 & \mbox{otherwise}
\end{array}
\right. .
$$
The creation (annihilation) operators anti-commute as ordinary fermions, whereas the operators 
satisfying 
Eq.(3.4) have no commutation relations at all. The algebra (3.17) is different from the algebra 
(3.4), but 
their corresponding statistics are the same.
The counting rule is given by Eq.(3.12).\\
The opposite example is
$$
\Theta_{jk}= \left\{ \begin{array}{ll}
1 &\mbox{if $ |k-j| \leq p$}\\
0 & \mbox{otherwise}
\end{array}
\right. 
$$
which implies that many-particle states satisfy the condition $N\leq p+1$ and all other states are 
forbidden.

\subsection{ Restricted Bose algebra}

Starting with the Bose algebra $a_i  a^{\dagger}_{j}= \delta_{ij} + a^{\dagger}_{j}a_i $, we can 
restrict it to 
a permutation invariant form 
\be
a_i  a^{\dagger}_{j}a^{\dagger}_{k}= (\delta_{ij} + a^{\dagger}_{j}a_i) 
\Theta_{jk}a^{\dagger}_{k},
\ee
where $\Theta_{jk}$ is given, for example,  as in subsection (3.4).\\
In the first example ($\Theta_{jk}=1,\, |k-j| \geq p$), the creation (annihilation) operators 
commute and the 
corresponding statistics are the same as for the quon-projected construction, Eqs.(3.4). The 
second example 
($\Theta_{jk}=1,\, |k-j| \leq p$) is equivalent with any of $(p+1)$- neighbouring Bose oscillators 
inside 
the initial lattice.\\
Generally, one can start with any algebra
$$
a_i  a^{\dagger}_{j}=\Gamma_{ij}(a^{\dagger};a)
$$
and restrict it in the following way
$$
a_i  a^{\dagger}_{j}a^{\dagger}_{k} =\Theta_{jk}\Gamma_{ij}(a^{\dagger};a)a^{\dagger}_{k}
$$
and proceed as in the above examples. 
We also note that the restricted Fermi and Bose statistics can 
be defined not only on a line, but also on a circle or a lattice with periodic boundary 
conditions. 
A special kind of this type of statistics with
$$
\Theta_{jk}= \left\{ \begin{array}{ll}
1 &\mbox{if $ |k-j| \geq p$}\\
0 & \mbox{otherwise}
\end{array}
\right. 
$$
is given in Ref.[26].\\

${\it Remarks}$

The construction of the C-S type fractional statistics and generalization proposed in subsections 
(3.2) and (3.3) 
( but not (3.4) and (3.5) ) relies crucially on the oriented 1-D space (lattice). However, we 
point out that this 
obstacle can be evaded and we suggest some interesting physical speculations.\\
If the creation (annihilation) operators are defined on $ {\cal{M}}_1 \otimes {\cal {M}}_2 $, 
where
 ${\cal{M}}_1$ is
D-dimensional and $ {\cal {M}}_2$ is 1-dimensional space, then one can apply the construction 
described in 
subsections (3.2) and (3.3). For example, there can be ordinary bosons and fermions in the $ {\cal 
{M}}_1$ direction but 
fractional in the $ {\cal {M}}_2$ direction:
$$
a_{i\alpha}  a^{\dagger}_{j\beta}a^{\dagger}_{k\gamma}=\delta_{\alpha \beta} (\delta_{ij}\pm 
a^{\dagger}_{j\beta}a_{i\alpha}) \Theta_{\beta \gamma}a^{\dagger}_{k\gamma}, 
$$
$$
\alpha , \beta , \gamma \in {\cal{M}}_2 \qquad i, j, k \in {\cal{M}}_1,
$$
where $\Theta_{\beta \gamma}$ is described in subsection (3.4). The whole space is not isotropic 
i.e.
there is a preferable direction ${\cal {M}}_2$. This is one of the main assumptions for the 
appearance 
of generalized statistics.

\section {Gentile - type statistics : Restrictions on each single oscillator}
\setcounter{equation}{0}
Gentile  suggested the first interpolation between Bose and Fermi statistics. 
It is characterized by the maximal occupation number  $m$  of particles (excitations) 
in a given quantum box. The maximal number of many-particle states is $N_{max}=m\, M$. 
The $m$ states available by one oscillator can be interpreted as internal degrees of 
freedom. For a single oscillator, 
$g_{n\rightarrow n+1}=  d^{(1)}_n - d^{(1)}_{n+1} = 0 $ if $n+1<m$ and 1 if $n+1= m$.
 Hence, $g^{Hald}=\bar{g}=1/m$.\\
 However, if there are M oscillators, the N-particle state is characterized by $1^{N_1}\cdots 
M^{N_M}$ 
 such that $\sum_{i=1}^{M}N_i = N$, $N_i\leq m$, where $i$ enumerates oscillators 1,2,...M. 
Alternatively, 
 we can write $0^{n_0}1^{n_1}....m^{n_m}$, such that $\sum_{\alpha =0}^m  n_{\alpha } =M$, 
 $\sum_{\alpha =0}^m \alpha n_{\alpha } = N$, where $n_{\alpha}$ denotes the number of oscillators 
with 
 $\alpha $-
 particles (excitations) . Let $N_{i_1} \geq N_{i_2}\geq \cdots \geq N_{i_n}$, then $n_{\alpha}$ 
is 
 number of boxes (oscillators) with $\alpha $ particles, $N_{k+1} = N_{k+2}= \cdots 
N_{k+n_{\alpha}}=\alpha $.
 Then $d^{(1)}_N=M-n_m$ and
 \be
 g_{N\rightarrow N+i}= \Delta n_m =1
 \ee
 if $"i"$ is added to the $(m-1) $ filling and 
 \be
 g_{N\rightarrow N+i}= \Delta n_m = 0
 \ee
 if $"i"$ is added to $n_{\alpha }$, $\alpha \leq m-2 $. We find
 \be
\bar{g}_{N \rightarrow N+1}=\frac{n_{m-1}}{M-n_m}.
\ee
Note that the average value is different from $1/m$, even if we perform averaging over different 
$N$ 
(except for the case of the single oscillator $M=1$).\\
Generally, the Gentile-type statistics can be defined by 
\be
a_i  a^{\dagger}_{j}=\Gamma_{ij}(a^{\dagger};a)\, \Theta (N,m) ,
\ee
where $\Theta (N,m) > 0$ for $N\leq m$ and $\Theta (m,m)=0$. The simplest functions with these 
properties are step-functions 
$\Theta(N-m)$ and $\Theta (N,m)=1-\frac{N}{m}$.

As example, 
we consider restricted Bose oscillator
\be
aa^{\dagger}=(1+a^{\dagger}a)\,\Theta(m-N),
\ee
with Fock space spanned by $|0>, a^{\dagger}|0>, ...(a^{\dagger})^m |0> $. This oscillator 
coincides 
with the truncated Bose oscillator with cutoff [22]
\be
aa^{\dagger}=(1+a^{\dagger}a) -(N+1)\, \delta_{N,m}
\ee
or $aa^{\dagger}=\Theta (m-N+1)$.
The counting rule is 
$$
D(N,M,m) = \sum_{\sum n_{\alpha} = M ; 
\sum\alpha  \, n_{\alpha }= N} 
( \frac{M!}{n_0!n_1!\cdots n_m! } )
$$
\be
D^F(M,N) \leq D(M,N,m)\leq D^B(M,N)
\ee

General properties of the Gentile-type algebra is that (i) extended Haldane parameters are not 
constant 
(ii) the average value of extended Haldane statistics parameters differs from $\frac{1}{m}$, $m 
\in {\bf N}$ 
except for a single oscillator for which $g_{n\rightarrow n+1} = \delta_{n,m}$ and 
$\bar{g}=\frac{1}{m}$ for 
 $n\leq m$ (iii) the counting rule differs from the Haldane-Wu formula for $m\neq 1$ and (iv) 
thermodynamic properties are 
 different from the Haldane-Wu thermodynamics [8]. These results are in agreement with the results 
obtained by 
 Chen et al. in Ref.[15].

\subsection {Karabali - Nair realization of Gentile statistics}
All algebras with the Gentile-type statistics satisfy $(a_i)^m \neq 0$, but $(a_i)^{m+1} = 0$ for 
every $i=1,2,\cdots M$, and the states $a^{\dagger}_ia^{\dagger}_j | 0 > $ and 
$a^{\dagger}_ja^{\dagger}_i | 0 > $, $i\neq j$, describe the same physical state. Karabali and 
Nair  constructed a special 
type of Gentile statistics in one dimension which is also of anyonic type.
The corresponding statistics has all properties of Gentile statistics and differs from the 
original 
Haldane statistics.\\
The simplest algebra with Gentile statistics in one dimension is of the form [17,27]
\be
a_ia^{\dagger} _j - e^{i\lambda sgn(i-j)}a^{\dagger}_j a_i  =0 \qquad \lambda = \pm \frac{2\pi 
}{m+1}, \quad m\in {\bf N}.
\ee
 For this 
algebra, we have $a_ia^{\dagger}_i=\Phi (N_i)$, $\Phi (n)= \frac{sin (n\lambda /2)}{sin (\lambda 
/2)} > 0$ 
for $n <m $ and  $\Phi (m+1)=0.$

{\it Remarks}\\
If the boxes are filled with small number of particles, after adding a few new particles , the 
system 
behaves like a Bose system. Contrary to this, if all boxes are filled with ($m-1$) particles the 
system behaves like 
a Fermi system. 
For Gentile statistics for large number of states is $g_{N\rightarrow N+i} = 0$ and for 
some N-particle states $g_{N\rightarrow N+i} > 0.$\\
 Finally, let us 
mention that besides  "local" restrictions described in this paper, there are "global" 
restrictions on  Fock space. Examples are Green's para-Bose and para-Fermi statistics [2]  of 
order $p$ , which can be 
realized through projections of complete quon Fock space [26] and parastatistics in which only 
states with $N\leq N_0$ particles are allowed [23], regardless of their local structure.


\section{Exclusions  on the base space   and single - oscillator Fock space }
\setcounter{equation}{0}
The construction of exclusion statistics performed in the preeceding sections can be 
combined to include restrictions between neighbours, as well as cutoff of single 
oscillators.We  present two examples of such exclusions, which 
includes  parastatistics introduced by Palev [23,28].
Consider the algebra
\be
a_ia^{\dagger}_ja^{\dagger}_k=f(N) (\delta_{ij} - a^{\dagger}_ja_i)a^{\dagger}_k,
\ee
with $f(n) >0$, $n < p$ and $f(p)=0$. The simplest choice is the step function 
$f(N)=\Theta(p-N)$  ( $\Theta (x) =0 , \, x \leq 0 $ and $ \Theta (x) = 1 , \, x>0 $ ).

 We point out that the corresponding statistics is Fermi statistics 
 restricted up to $N\leq p$ N-particle states. Hence, the counting rule is simply 
 $D^F(M,N)=\left( \begin{array}{c}
M\\ N
\end{array} \right) $, $N\leq p $ and  $D^F(M,N)=0$ if $N > p $.
The above statistics is characterized by the Haldane statistical parameter $g=1$

\be
g_{n\rightarrow n+k}=\frac{d_n -d_{n +k}}{k}=\frac{(M-n+1)-(M-n-k+1)}{k} =1 ,
\ee
if $n+k\leq p$. If $n+k = p+1$, then $g_{n\rightarrow n+k}=\frac{(M-n+1)}{(p-n+1)}$, 
$n=1,2,\cdots p $ is 
fractional but $g$ is not constant any more. Hence, this is not an example for the 
original Haldane statistics for which the statistics parameter is $g = const.$ Moreover, the 
above statistics is also not the statistics of the Karabali-Nair type , where $a_i^p \neq 0$, 
$a_i^{p+1}=0$, 
and for any $N\leq M p$ N-particle state is allowed, since  we already have 
$a_i^2=0$ and $N\leq p$.

The second example is the Bose counterpart of the algebra (5.1 ), namely:

\be
a_ia^{\dagger}_ja^{\dagger}_k=f(N) (\delta_{ij} + a^{\dagger}_ja_i)a^{\dagger}_k,
\ee
with $f(n) >0$, $n < p$ and $f(p)=0$. The simplest choice is the step function mentioned after 
Eq.(5.1)
 or
$f(N)= 1 - \frac{N}{p}$.
 The corresponding statistics is Bose statistics 
 restricted  to  N-particle states with $N\leq p$. Hence, the counting rule is simply 
 $D^B(M,N)=\left( \begin{array}{c}
M+N-1\\ N
\end{array} \right)$, $N\leq p $ and  $D^B(M,N)=0$ if $N > p $. Therefore, the above 
statistics is characterized by the Haldane statistics parameter $g=0$
\be
g_{n\rightarrow n+k}=\frac{d_n -d_{n +k}}{k}=\frac{M - M}{k} =0 ,
\ee
if $n+k \leq p$. If $n+k = p+1$, then $g_{n\rightarrow n+k}=\frac{M}{(p-n+1)}$, 
$n=1,2,\cdots p$, is 
fractional but  not constant . Hence, this is not an example for the 
original Haldane exclusion statistics.  The 
above statistics is also not  of the Karabali-Nair type , since $a_i^p \neq 0$, 
 $a_i^{p+1}=0$ but $N\leq p$. This would be equivalent only for the single-mode oscillator, $M=1$.

Let us mention that Green's para-Fermi statistics [2] of order ${\it p} \in \bf N $
is also example of this kind of exclusion statistics since at most ${\it p}$ particles can occupy 
a given 
quantum state, $a_i^{p+1}=0$. For a single oscillator $a^{p+1}=0$, the extended statistical 
parameters are  $ g_{i\rightarrow j}= 0 $ for  $j\leq p$ and 
$g_{i\rightarrow p+1}=\frac{1}{p+1-i}$ .\\
In a recent papers [16,28], we have discussed these algebras and statistics in more details.


\section{Summary}
In  previous papers [16,28], we defined the extended Haldane statistics parameters 
${\it g}$, ( see Eq.(2.6)) and the counting rules $D(M,N;\Gamma)$ ( see Eq.(2.5)),
for the generalized statistics formulated in the second quantized approach.\\
In this paper, we have proposed and further analyzed three types of 
exclusion statistics, namely, the Calogero-Sutherland (C-S) type, the Gentile type and 
a combination of these two types of exclusion statistics.\\
We have started with the multimode oscillator algebra $\Gamma_{ij}$, Eq.(2.3), with positive 
definite 
Fock space.  Introducing the appropriately defined step - function 
$\Theta_{jk}$, we have restricted the algebra $\Gamma_{ij}$ in such a way that certain states in 
the 
Fock space of the algebra $\Gamma_{ij}$ are forbidden, i.e. they have zero norms by 
construction.\\
The realization of the C-S type of exclusion statistics relies on the quon algebra, Eq.(3.3), and 
on the restriction between neighbour oscillators (  placed on the 1D lattice ), 
induced by the step-function, Eq.(3.4). For this type of statistics, we have calculated extended
statistics parameters  ${\it g}$ for the finite lattice, Eq.(3.7), and the infinite lattice,
Eq.(3.9). We have also defined and calculated the average value of the extended statistics 
parameters, 
Eq.(3.8). Furthermore, we have calculated counting rule, Eq.(3.10), and discussed its dependence 
on the 
parameters $p$ and $q$, Eqs.(3.11-3.13). In the subsections (3.2-3.5), we have briefly described 
the possible 
generalization of the above procedure.\\
As an example of the Gentile type of exclusion statistics, we have considered the  Bose-like 
algebra with 
$(a^{\dagger})^{m+1}=0$ and the step-function $\Theta (m-N)$, Eq.(4.5). This is a restriction 
in the single-oscillator Fock space. We have found that the extended statistics parameters are not 
constant and that the counting rule differs from the Haldane-Wu formula. We have also mentioned 
the Karabali-Nair realization of the Gentile statistics.\\
Finally, we have described the combination of these two exclusions. As an example, we discussed 
parastatistics, Eqs.(5.1,5.2). We have found that  the extended statistics parameters are 
fractional 
but not constant  .\\
None of the examples of exclusion presented here, include the original Haldane proposal [5].
 As we stressed before [16], it seems that the original Haldane statistics cannot be realized 
 in the above sense, i.e. one cannot define the underlying operator algebra of creation and 
 annihilation operators with positive Fock space and satsfying Haldane's requirements 
 (  fractional and constant $g$ ), 
 ${\it except}$ for ${\it free}$ bosons and fermions. One should recall that the 
 Haldane fractional exclusion statistics arises because the system is an 
 ${\it interacting}$ system and  particles are topological excitations of a 
 condensed matter state, rather than real particles which can exist outside the finite 
 region of condensed matter.
 Hence, our analysis confirms Haldane statement that the techniques of the sceond-quantized 
 many-body theory 
 cannot be applied to this type of exclusion statistics.


\bigskip

{\bf Acknowledgment}
The authors would like to thank D.Svrtan for useful discussions. We also thank 
 referees for useful remarks.


\newpage
\baselineskip=24pt
{\bf References}
\begin{description}

\item{[1]}
Gentile G 1940 N.Cimento {\bf 17} 493 
\item{[2]}
Green H S 1953 Phys.Rev.{\bf 90} 170 \\
 Greenberg O W and Messiah A M L  
 1965 Phys.Rev.B {\bf 138} 1155 ; J.Math.Phys. {\bf 6} 500 \\
Ohnuki Y and Kamefuchi S 1982 {\it Quantum Field Theory and Parastatistics} 
( University of Tokio Press, Tokio, Springer, Berlin, 1982).
\item{[3]}
Prange R E and Girvin S M (eds) 1990 {\it The Quantum Hall Effects }
(Springer,Berlin 1990)\\
Stone M(ed.) 1992 {\it  Quantum Hall Effect(World Scientific },
Singapore,1992)\\
Canright G S and Johnson M D 1994 J. Phys.A :Math.Gen.{\bf 27} 3579 
\item{[4]}
Wilczek F 1990 {\it Fractional statistics and anyon superconductivity } (World Scientific,
Singapore,1990)
\item{[5]}
Haldane F D M 1991 Phys.Rev.Lett. {\bf 67} 937 
\item{[6]}
Greenberg O W 1990 Phys.Rev.Lett. {\bf 64} 705 \\
Greenberg O W 1991 Phys.Rev. D{\bf 43 } 4111 \\
Meljanac S and Perica A  1994 Mod.Phys.Lett.A {\bf 9}  3293 
\item{[7]}
Wu Y S  1994 Phys.Rev.Lett. {\bf 73} 922 
\item{[8]}
Protogenov A P and Verbus V A  1997 Mod.Phys.Lett.B {\bf 11 } 283 \\
Ho C L and Liao M J 1997  Mod.Phys.Lett.B {\bf 11} 461 \\
 Iguchi K 1997 Phys.Rev.Lett. {\bf 78 } 3233 ; Mod.Phys.Lett.B {\bf 11 } 765 \\
Chaturvedi S and Srinivasan V  1997 Phys.Rev.Lett. {\bf 78 } 4316\\
Isakov S, Arovas D, Myrheim J and  Polychronakos A P  1996 Phys.Lett.A {\bf 212 } 299
 \\
  Polychronakos A P 1996 Phys.Lett.B {\bf 365} 202 \\ Bhaduri R K, Murthy M V N and Srivastava M K 
1996
 Phys.Rev.Lett. {\bf 76 } 165 \\
 Rajagopal A G Phys.Rev.Lett. 1995 {\bf 74 } 1048 \\
 Fukui T and Kawakami N  1995 Phys.Rev.B {\bf 51 } 5239 \\ Sen D and Bhaduri R K 1995
Phys.Rev.Lett. {\bf 74 } 3912 \\
Isakov S 1994 Mod.Phys.Lett.B {\bf 8} 319  
\item{[9]}
He S, Xie X C and Zhang F C  1992 Phys.Rev.Lett. {\bf 68 } 3460  \\
Johnson M D and Canright G S  1994 Phys.Rev.B{\bf 49 } 2947 \\
Li D and Ouvry S 1994 Nucl.Phys.B {\bf 430 [FS]} 563 \\
Su W P , Wu Y S and Yang J 1996 Phys.Rev.Lett. {\bf 77 } 3423 \\
Isakov S, Canright G S and Johnson M D  1997 Phys.Rev.B{\bf 55 } 6727 
\item{[10]}
Dasnieres de Veigy A and Ouvry S  1994 Phys.Rev.Lett. {\bf 72} 600 \\
Dasnieres de Veigy A and Ouvry S 1995 Mod.Phys.Lett.A {\bf 10} 1 ; Mod.Phys.Lett.B {\bf 9} 271 \\
Murthy M V N and Shankar R 1994 Phys.Rev.Lett. {\bf 72} 3629 \\
Chen W and Ng Y J  1995 Phys.Rev.B{\bf 51 } 14479 \\
Fayyazuddin A and Li D 1996 Phys.Rev.Lett. {\bf 76} 1707 \\
Isakov S and Mashkevich S 1997 Nucl.Phys.B {\bf 504 [FS]} 701
\item{[11]}
Bernard D and Wu Y S 1994 in  {\it New Developments of Integrable Systems and 
Long-Ranged Interaction Models } (Nankai Lectures on Mathematical Physics,
 World Scientific, Singapore 1994)\\
 Isakov S  1994 Int.J.Mod.Phys. A. {\bf 9} 2563 \\
 Murthy M V N and Shankar R 1994 Phys.Rev.Lett. {\bf 73} 3331 \\
Ha Z N C Phys.Rev.Lett. 1994 {\bf 73} 1574\\
Dasnieres de Veigy A and Ouvry S 1995 Phys.Rev.Lett. {\bf 75} 352 \\
Ujino H and Wadati M  1995 J.Phys.Soc.Jpn. {\bf 64} 4064 \\
Ujino H and Wadati M  1996 J.Phys.Soc.Jpn. {\bf 65} 1203 \\
 Mashkevich S 1997 Phys.Lett.A, {\bf 233} 30 
\item{[12]}
Ha Z N C 1995 Nucl.Phys.B {\bf 435 [FS]} 604 
\item{[13]}
Wu Y S and Yu Y 1995 Phys.Rev.Lett. {\bf 75} 890 
\item{[14]}
Nayak C and Wilczek F 1994 Phys.Rev.Lett. {\bf 73} 2740\\
Hatsugai Y, Kohmoto M, Koma T and Wu Y S 1996 Phys.Rev.B {\bf 54 } 5358 
\item{[15]}
 Karabali D and Nair V P 1995
Nucl.Phys.B {\bf 438 [FS]} 551 \\  Chen W,  Ng Y J and  Van Dam H 1996
Mod.Phys.Lett.A {\bf 11} 795 \\ Speliotopoulos A 1997
J. Phys.A :Math.Gen.{\bf  30 } 6177 
\item{[16]}
Meljanac S and Milekovic M  1996 Mod.Phys.Lett.A {\bf 11} 3081 
\item{[17]}
Meljanac S and Milekovic M  1996 Int.J.Mod.Phys.A. {\bf 11} 1391 \\
 Melic B and Meljanac S 1997 Phys.Lett.A {\bf 226} 22 \\
Meljanac S, Stojic M and Svrtan D 1997 Phys.Lett.A {\bf 224 } 319 
\item{[18]}
Jorgensen P, Schmitt L and Werner R 1995 J.Funct.Anal. {\bf 134 } 33\\
Bozejko M and Speicher R 1994 Math.Ann. {\bf 300 } 97
\item{[19]}
Zagier D 1992 Comm.Math.Phys.{\bf 147} 199 \\
Meljanac S and Svrtan D 1996  Comm.Math. {\bf 1} 1 
\item{[20]}
Chou C 1992 Mod.Phys.Lett.A  {\bf 7} 2685 \\ 
 Debergh N 1993
Mod.Phys.Lett.A {\bf 8} 765 
\item{[21]}
Meljanac S, Milekovic M and Pallua S 1994 Phys.Lett.B {\bf 328} 55 \\
Bonatsos D and Daskaloyannis C  1993 Phys.Lett.A{\bf 8} 3727 
\item{[22]}
Jevicki A and van Tonder A 1996 Mod.Phys.Lett. A {\bf 11} 1397 
\item{[23]}
Palev T  1982 J.Math.Phys.{\bf 23} 1778 \\
Palev T and Stoilova S  1994 J.Phys.A :Math.Gen.{\bf 27} 977 
ibid. {\bf 27} 7387 \\
Palev T and Stoilova S 1997 J.Math.Phys.{\bf 38} 2806 \\
Okubo S 1994 J.Math.Phys.{\bf 35} 2785 
\item{[24]}
Aref'eva I Ya and Volovich I V 1996 Nucl.Phys.B {\bf 462} 600 
\item{[25]}
Goulden I P and Jackson D M 1983 {\it Combinatorial Enumeration} (J.Wiley ,1983).
\item{[26]}
Polychronakos A P 1995 Phys.Lett.B {\bf 365} 202 
\item{[27]}
Meljanac S, Milekovic M and Perica A 1994 Europhys.Lett. {\bf 28} 79 \\
Doresic M, Meljanac S and Milekovic M 1994 Fizika B{\bf 3} 57 
\item{[28]}
Meljanac S, Milekovic M and Stojic M 1998 Mod.Phys.Lett.A {\bf 13} 995

\end{description}
\end{document}